\documentclass[useAMS,usedcolumn,usenatbib]{mn2e}
\usepackage{graphicx,color}
\usepackage{url}
\usepackage[normalem]{ulem} 

\usepackage{times,amsmath}

\newcommand{\alm}{a_{\ell m}}
\newcommand{\pseudoalm}{\tilde{a}_{\ell m}}
\newcommand{\Cee}{\mathcal{C}}
\newcommand{\Cl}{\Cee_\ell}
\newcommand{\pseudoCl}{\tilde\Cl}
\newcommand{\pseudoClest}{\Cl^{\mathrm{cut}}}
\newcommand{\Ccorr}{\Cee} 
\newcommand{\pseudoCcorr}{\Ccorr^{\mathrm{cut}}}

\newcommand{\ellmax}{\ell_{\rmn{max}}}
\newcommand{\fsky}{f_{\rmn{sky}}}
\newcommand{\Nside}{\textit{Nside}}

\newcommand{\muK}{\rmn{\umu K}}
\newcommand{\spice}{\textsc{spice}}
\newcommand{\healpix}{\textsc{healpix}}

\newcommand*{\satellite}[1]{\textit{#1}}
\newcommand{\COBE}{\satellite{COBE}}
\newcommand{\COBEDMR}{\satellite{COBE-DMR}}
\newcommand{\WMAP}{\satellite{WMAP}}
\newcommand{\Planck}{\satellite{Planck}}
\newcommand{\LFI}{\satellite{LFI}}
\newcommand{\HFI}{\satellite{HFI}}

\newcommand*{\Planckmap}[1]{\texttt{#1}}
\newcommand{\smica}{\Planckmap{SMICA}}
\newcommand{\nilc}{\Planckmap{NILC}}
\newcommand{\sevem}{\Planckmap{SEVEM}}
\newcommand{\CR}{\Planckmap{Commander-Ruler}}

\newcommand{\LCDM}{$\Lambda$CDM}

\newcommand*{\unit}[1]{\;\rmn{#1}}
\renewcommand*{\vec}[1]{\bmath{#1}}
\newcommand*{\unitvec}[1]{\vec{\hat{#1}}}

\newcommand{\Shalf}{\ensuremath{S_{1/2}}}

\newcommand{\dderiv}{\rmn{d}}

\title[Lack of large-angle TT correlations persist]{Lack of large-angle TT
  correlations persists in WMAP and Planck}
\author[C.J. Copi, D. Huterer, D.J. Schwarz and G.D. Starkman]
{Craig J. Copi$^{1}$\thanks{E-mail: cjc5@cwru.edu},
 Dragan Huterer$^{2}$\thanks{E-mail: huterer@umich.edu},
 Dominik J. Schwarz$^{3}$\thanks{E-mail: dschwarz@physik.uni-bielefeld.de}
 and 
 Glenn D. Starkman$^{1,4}$\thanks{E-mail: glenn.starkman@case.edu}\\
 $^{1}$CERCA/Department of Physics/ISO, Case Western Reserve University,
 Cleveland, OH 44106-7079, USA\\
 $^{2}$Department of Physics, University of Michigan, 
 450 Church St, Ann Arbor, MI 48109-1040, USA\\
 $^{3}$Fakult\"at f\"ur Physik, Universit\"at Bielefeld,
 Postfach 100131, 33501 Bielefeld, Germany\\
 $^{4}$Physics Department, Theory Unit, CERN,
 CH-1211 Gen\`eve 23, Switzerland}

\begin{document}

\date{Accepted xxxx. Received xxxx; in original form xxxx}

\pagerange{\pageref{firstpage}--\pageref{lastpage}} \pubyear{2013}

\maketitle

\label{firstpage}

\begin{abstract}
  The lack of large-angle correlations in the observed microwave background
  temperature fluctuations persists in the final-year maps from \WMAP\ and
  the first cosmological data release from \Planck. We find a statistically
  robust and significant result: $p$-values for the missing correlations
  lying below $0.24$ per cent (i.e.\ evidence at more than $3 \sigma$) for
  foreground cleaned maps, in complete agreement with previous analyses
  based upon earlier \WMAP\ data.  A cut-sky analysis of the
  \Planck\ \HFI\ $100\unit{GHz}$ frequency band, the `cleanest CMB channel'
  of this instrument, returns a $p$-value as small as $0.03$ per cent,
  based on the conservative mask defined by \WMAP\@.  These findings are in
  stark contrast to expectations from the inflationary Lambda cold dark
  matter model and still lack a convincing explanation.  If this lack of
  large-angle correlations is a true feature of our Universe, and not just
  a statistical fluke, then the cosmological dipole must be considerably
  smaller than that predicted in the best-fitting model.
\end{abstract}


\section{Introduction}

The first release of cosmological data from the \Planck\
satellite \citep{Planck-R1-I} and the final analysis of the
\satellite{Wilkinson Microwave Anisotropy Probe} (\WMAP)
\citep{WMAP9-results} confirmed that the inflationary Lambda cold dark
matter (\LCDM) model provides an excellent fit to the angular temperature power
spectrum for multipoles ranging from the quadrupole ($\ell = 2$) up to
$\ell = 2500$. The effect of gravitational lensing of the cosmic microwave
background (CMB) has been detected with a very high statistical
significance ($25 \sigma$) \citep{Planck-R1-XVII} and breaks some parameter
degeneracies without reference to non-CMB observations. Most of the
statistical power in the \Planck\ analysis comes from high-$\ell$
multipoles, thus it may not come as a surprise that the best-fitting model
traces the high-$\ell$ data much better than those at low-$\ell$, where a
lack of angular power (in the range $\ell = 2$ to $32$) compared to the
best-fitting model is found at the $99$ per cent C.L.\ \citep{Planck-R1-XV}.
Nevertheless, it is quite remarkable that none of the models invoking
additional, physically well motivated parameters, such as the sum of
neutrino masses, the number of effective relativistic degrees of freedom,
or a running of the spectral index, can give rise to a significant
improvement of the fit \citep{Planck-R1-XVI}.  These findings indicate that
some special attention should be devoted to the largest angular scales,
especially as they potentially probe different physics than the small angular
scales.

Several anomalies at large angular scales discussed in the literature have
been the source of some controversy since the first release of the
\WMAP\ data (see \citealt{WMAP7-anomalies,CHSS-review} for reviews). The
first of them was already seen by the \satellite{Cosmic Background
  Explorer} (\COBE): the temperature two-point angular correlation function
computed as an average over the complete sky
\begin{equation}
  \Ccorr(\theta) = \overline{T(\unitvec e_1) T(\unitvec e_2)}, \qquad
  \unitvec e_1 \cdot \unitvec e_2 = \cos \theta, 
  \label{eq:Ctheta-full-sky}
\end{equation}
was found to be smaller than expected at large angular scales
\citep{DMR4-Ctheta}. Scant attention was given to this observation, due in
part to the relatively low signal to noise ratio of the
\COBE\ observations, but mostly to the theory-driven shift in attention
away from the angular correlation function and toward the angular power
spectrum.  The lack of correlations on angular scales larger than $60$
degrees was rediscovered almost a decade later by \WMAP\ in their one-year
analysis \citep{WMAP1-cosmology} and analysed in greater detail by us for
the \WMAP\ three and five-year data releases \citep{CHSS-WMAP5}.  We have
emphasized its persistence in the data (contrary to some claims),
differentiated it from the lowness of the temperature quadrupole with which
it is often confused, and demonstrated how it challenges the canonical
theory's fundamental prediction of Gaussian random, statistically isotropic
temperature fluctuations.  For related work on the missing large-angle
correlations, see also
\cite{CHSS-WMAP3,Hajian2007,SHCSS2011,Kim2011,Zhang2012,Gruppuso2014}.

The \Planck\ team presented an analysis of the angular two-point correlation
function at a low resolution ($\Nside = 64$) for their four component
separation methods (\CR, \nilc, \sevem, \smica) after the U73 mask was used to
suppress Galactic residuals.  Based on comparison with $10^3$ realizations of
the best-fitting model, they find that the probability of obtaining a $\chi^2$
between the expected angular two-point correlation function of the
best-fitting model and the observed correlation function that is at least as
large as that measured is $0.883$, $0.859$, $0.884$, and $0.855$ for the \CR,
\nilc, \sevem, and \smica\ maps respectively \citep{Planck-R1-XXIII}. However,
their statistic fails to capture that what is anomalous about the angular
two-point correlation function is not the extent to which it deviates from the
theoretical expected value of the function.  Rather,  as has been the
  case since the \COBEDMR\ observation, the pertinent anomaly is
that above about $60$ degrees the angular correlation function is very nearly
zero.  It is this very special way of deviating from our expectation that
  deserves our attention.

In this work, we analyse the two-point angular correlation function at
large angles as seen in the final data release of \WMAP\ and the first
cosmology release of \Planck.  The anomalous alignments of low multipole
modes with each other and with directions defined by the geometry and
motion of the Solar system are discussed in a companion paper
\citep{CHSS-Planck-R1-alignments}.  Here we demonstrate that on the part of
the sky outside the plane of the Galaxy the absence of two-point angular
correlations above about $60$ degrees remains a robust, statistically
significant result, with a $p$-value between about $0.03$ and $0.33$ per
cent depending on the precise map and Galaxy cut being analysed.

\section{Physics at large angular scales}

High fidelity measurements of the microwave sky reveal the imprints
of primary temperature, density and metric fluctuations in the early
Universe.  By observing these fluctuations and analysing their statistical
properties, we seek a deeper understanding of cosmological inflation or any
alternative mechanism that produced the initial fluctuations.

Studying modes with wavelengths too large to enable causal contact across
the mode during the radiation and most of the matter dominated epochs
suggests that we can learn something about the physics of inflation without
detailed knowledge of the recent content of the Universe and associated
astrophysical details (e.g.\ reionization).  This motivates us to pay
special attention to the largest angular scales.  Comoving scales that
cross into the Hubble radius at $z \sim1$ and below are observed at angles
larger than $60$ degrees. Thus features observed at those scales are either
of primordial nature or stem from physics at $z \la 1$, the epoch in the
history of the Universe that we arguably know best.

To be more precise, at $z = 0.91 (1.5, 7)$ the comoving Hubble length
equals the length of a comoving arc with an opening angle of $90\degr
(60\degr,18\degr)$ for the best-fitting \LCDM\ model. These angular
scales correspond roughly to the scales that have been shown to be
anomalous in previous works (the quadrupole, octopole, and up to modes $\ell =
10$). It is possibly  noteworthy that $z \sim 7$ corresponds to the moment
when the Universe is fully reionized.

For better or worse, however, the large-angle CMB is also sensitive to the
physics that affects the microwave photons as they propagate from their
last scattering until their collection by our telescopes.  The late-time
integrated Sachs-Wolfe (ISW) effect could potentially correlate the
large-angle CMB with the local structure of the gravitational
potential. Indeed, it has been proposed in the literature that some of the
observed CMB anomalies could be explained in this way
\citep{Rakic2006a,Francis2010,Dupe2011,Rassat2013}.  Although
reconstruction of the local gravitational potential from existing CMB and
large-scale structure data is quite uncertain and subject to biases, such
an explanation would indeed be an attractive possibility if only there were
no lack of correlations on large scales.
If the observed lack of large-angle correlations is real, then we must
explain how the local gravitational potential manages to align with the
primordial temperature fluctuations in such a way that the resulting sky
has such a deficit.  In the end this does not change the underlying
problem; it merely rephrases it from one about the CMB to one about the
local gravitational potential.
  
Clearly, it is important to understand the lack of correlations at large
angular scales in greater detail not just for its own sake, but also in
order to evaluate any proposed explanation for other features of the CMB
data, especially other large-angle or low-$\ell$ anomalies.

\section{Temperature two-point angular correlation function} 

\subsection{Theory}

In the standard CMB analysis a full-sky map of temperature fluctuations,
$T(\unitvec e)$, is expanded in spherical harmonics as
\begin{equation}
  T(\unitvec e) = \sum_{\ell m} \alm Y_{\ell m}(\unitvec e),
\end{equation}
where the coefficients in the expansion are extracted from the full-sky as
\begin{equation}
  \alm = \int T(\unitvec e) Y_{\ell m}^*(\unitvec e)\, \dderiv\unitvec e.
\end{equation}
From these quantities we define the angular power spectrum as
\begin{equation}
  \Cl \equiv \frac1{2\ell+1} \sum_{m}|\alm|^2.
  \label{eq:Cl-full-sky}
\end{equation}
Note that the angular power spectrum may \emph{always} be defined in this
way.  Only in the case of Gaussian random, statistically isotropic
temperature fluctuations will it contain \emph{all} the statistical
information.  The full-sky two-point angular correlation
function~(\ref{eq:Ctheta-full-sky}) is related to the full-sky angular
power spectrum via a Legendre series
\begin{equation}
  \Ccorr(\theta) = \sum_\ell \frac{2\ell+1}{4\upi} \Cl P_\ell(\cos\theta),
  \label{eq:Ctheta-full-sky-expansion}
\end{equation}
where the $P_\ell(\cos\theta)$ are Legendre polynomials.

Unfortunately the full-sky cannot be observed due to foreground
contamination.  If we let $W(\unitvec e)$ represent a mask on the sky (in
the simplest case it is zero for pixels removed and one for those included)
then cut-sky quantities can be defined in analogy to the full-sky ones from
above.  In particular, the cut-sky two-point angular correlation function
is defined as the sky average,
\begin{align}
  \pseudoCcorr (\theta) & \equiv
  \overline{W(\unitvec e_1)T(\unitvec e_1) W(\unitvec e_2)T(\unitvec e_2)}
  \label{eq:Ctheta-cut-sky}
  \\
  & \equiv
  \frac{\sum_{ij} W(\unitvec e_i)T(\unitvec e_i) W(\unitvec e_j)T(\unitvec
    e_j)}{\sum_{ij} W(\unitvec e_i) W(\unitvec e_j)},
  \quad \unitvec e_i \cdot \unitvec e_j = \cos \theta,
  \nonumber
\end{align}
where the sums are over all pairs of pixels separated by an angle $\theta$.
This correlation function can be evaluated in harmonic space by first
expanding the cut-sky in pseudo-$\alm$ as
\begin{equation}
  W(\unitvec e) T(\unitvec e) = \sum_{\ell m} \pseudoalm Y_{\ell
    m}(\unitvec e),
\end{equation}
where
\begin{equation}
  \pseudoalm = \int W(\unitvec e) T(\unitvec e) Y_{\ell m}^*(\unitvec e)\,
  \dderiv\unitvec e.
\end{equation}
From these the pseudo-$\Cl$ are defined by
\begin{equation}
  \pseudoCl \equiv \frac1{2\ell+1} \sum_m |\pseudoalm|^2.
  \label{eq:Cl-cut-sky}
\end{equation}
Following \cite{polspice} it can be shown that
\begin{equation}
  \pseudoCcorr(\theta) = 2\upi A(\theta) \sum_\ell (2\ell+1) \pseudoCl
  P_\ell(\cos\theta).
  \label{eq:Ctheta-cut-sky-sum}
\end{equation}
Here the normalization, $A(\theta)$, depends on the mask and may be
calculated in harmonic space as
\begin{equation}
  \frac1{A(\theta)} = 2\upi \sum_\ell (2\ell+1) w_\ell P_\ell(\cos\theta),
\end{equation}
where
\begin{equation}
  w_\ell \equiv \frac1{2\ell+1} \sum_m |w_{\ell m}|^2
\end{equation}
and the $w_{\ell m}$ are coefficients from the spherical harmonic expansion
of the mask,
\begin{equation}
  W(\unitvec e) = \sum_{\ell m} w_{\ell m} Y_{\ell m}(\unitvec e).
\end{equation}
Notice that for the full-sky $w_{\ell m}=\sqrt{4\pi} \delta_{\ell 0}
\delta_{m 0}$ so that $2\upi A(\theta) = 1/4\upi$ and the cut-sky
expansion~(\ref{eq:Ctheta-cut-sky-sum}) reproduces the full-sky
result~(\ref{eq:Ctheta-full-sky-expansion}), as it must. Finally, since
$\pseudoCcorr(\theta)$ is a function defined on the interval
$-1\le\cos\theta\le 1$ it may be expanded in a Legendre series as
\begin{equation}
  \pseudoCcorr(\theta) = \sum_\ell \frac{2\ell+1}{4\upi} \pseudoClest
  P_\ell(\cos\theta).
  \label{eq:Ctheta-cut-sky-expansion}
\end{equation}
Note that it is common to just refer to $\Ccorr(\theta)$ as a single
quantity covering both the full- and cut-sky cases.  It should be
remembered that whenever a cut-sky $\Ccorr(\theta)$ is discussed it is
defined as in Eq.~(\ref{eq:Ctheta-cut-sky}) and it may be expanded in a
Legendre series using the cut-sky $\Cl$ as in
Eq.~(\ref{eq:Ctheta-cut-sky-expansion}).

For a statistically isotropic universe the ensemble average of the
pseudo-$\Cl$~(\ref{eq:Cl-cut-sky}) is related to the ensemble average of
$\pseudoClest$ through a mode coupling matrix \citep{Hauser1973} and
$\pseudoClest$ provides an unbiased estimator of the theoretical (full-sky)
angular power spectrum~(\ref{eq:Cl-full-sky}).  Lacking statistical
isotropy or some other model the cut-sky angular power spectrum,
$\pseudoClest$, can still be related to the pseudo-$\Cl$ through the same
mode coupling matrix (see \citealt{Pontzen2010} for a proof of this result)
however the utility of the $\pseudoCl$ or $\pseudoClest$ as estimators of
the full-sky or theoretical angular power spectrum would be completely
unknown in that case.

It should emphasized that the mathematical connection between cut-sky
quantities, $\pseudoCcorr$ and $\pseudoClest$, and $\pseudoCl$ does
\emph{not} rely on assumptions from a theory.  However, when measured
quantities are to be related to the properties of the ensemble predicted by
a theory assumptions such as Gaussianity and statistical isotropy become
important and must be identified.  Thus, to construct an estimator of the
theoretical angular power spectrum from cut-sky observations -- either
through the pseudo-$\Cl$ or a maximum likelihood technique -- extra
assumptions are required. These assumptions may not be valid on
large-scales (or low-$\ell$) even if they work well on small-scales (or
high-$\ell$).

The simple point being made here is that masking removes information from a
map.  Without assumptions regarding the properties of this information it
cannot be reinserted when a full-sky map is created.  Not even the
statistical properties of this information can be known without extra
assumptions.  In fact, the need for masking of a CMB map is precisely due
to contaminations in some regions of the sky. These contaminated regions
are excised from the map so as to not affect deduced properties of the
underlying theory.  Without assumptions a \emph{unique reconstruction of a
  full-sky map (or any quantity relying on the properties of the masked
  regions) cannot be computed sensibly}.  Particular assumptions may be
reasonable or expected to be valid, regardless, such assumptions are
required if the full-sky is to be reconstructed and are not required when
working solely with cut-sky quantities.  For this reason the cut-sky
two-point angular correlation function will be the sole focus of this work.

At high-$\ell$, observed deviations from Gaussianity agree with the amount
of non-Gaussianity expected from the non-linear contributions of
gravitational lensing \citep{Planck-R1-XXIV}.  However, at low-$\ell$,
there are statistically significant anomalies in the temperature map, such
as the alignments of multipoles and the hemispherical power asymmetry
\citep{CHSS-Planck-R1-alignments,Planck-R1-XXIII}, that are evidence of
correlations among the $\alm$ (for different values of $\ell$ and $m$), and
thus contradict the assumption of Gaussian-random statistically isotropic
$\alm$.  This suggests that the physics underlying the observed sky cannot
be characterized solely by the $\Cl$, the statistical quantities prescribed
by the canonical model; unless these anomalies are unfortunate `flukes',
other statistical tools are not just interesting but necessary.  The
difficultly comes in identifying which are the appropriate ones.  The
resolution clearly depends on the physics underlying the anomalies.  At
least until that physics is established, multiple approaches will need to
be explored.

\subsection{Analysis of Observations}
\begin{figure}
  \includegraphics[width=\linewidth]{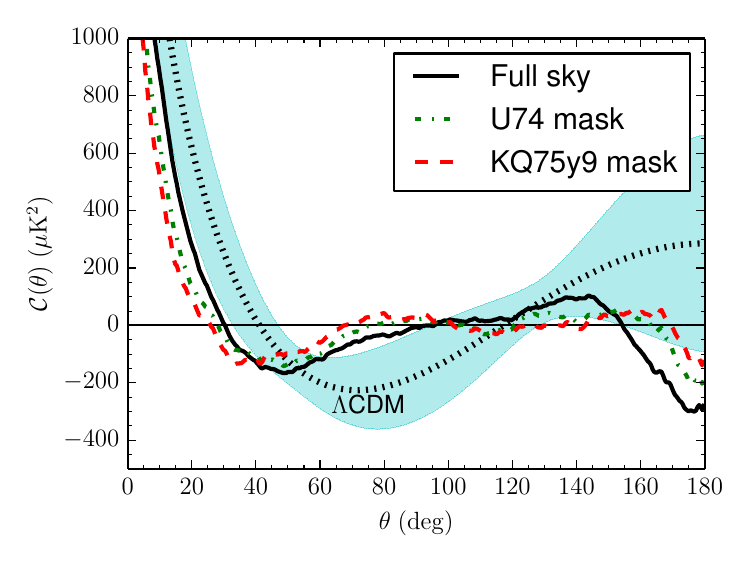}
  \caption{Two-point angular correlation function from the inpainted
    \Planck\ \smica\ map.  The black, dotted line shows the best-fitting
    \LCDM\ model from \Planck.  The shaded, cyan region is the $68$ per
    cent cosmic variance confidence interval.  Included from the
    \smica\ map are the $\Ccorr(\theta)$ calculated on the full-sky (black,
    solid line) and from two cut skies using the U74 mask (green,
    dash-dotted line) and the KQ75y9 mask (red, dashed line).  See the text
    for details.}
  \label{fig:Ctheta-smica}
\end{figure}

The two-point temperature angular correlation function for the CMB,
$\Ccorr^{TT}(\theta)$, has remained mostly unchanged since first measured
by the \COBEDMR\ \citep{DMR4-Ctheta}.  The resulting curves from the
\Planck\ \smica\ map are shown in Fig.~\ref{fig:Ctheta-smica}.  What is
most striking at first glance may be the difference between the
best-fitting \LCDM\ model and the observed $\Ccorr(\theta)$ on both the
full and cut skies (the details of the masks will be discussed below).
This is a source of considerable confusion and great care must be taken to
not read too much into this.  The values of $\Ccorr(\theta)$ at different
angular separations $\theta$ (or more precisely in different angular bins)
are correlated, so the sizeable deviation between the expected \LCDM\ and
the observed curves is not as significant as it may appear.  Rather, it is
the very small value of the observed $\Ccorr(\theta)$ on large angular
scales that is truly surprising.  This is particularly true for the cut
skies where there are essentially no correlations above about 60 degrees,
except for some small anti-correlation near 180 degrees.

To quantify this lack of correlations on large angular scales we continue to
use the statistic first proposed in the \WMAP\ one-year analysis
\citep{WMAP1-cosmology},
\begin{equation}
  \label{eq:Shalf}
  \Shalf \equiv \int_{-1}^{1/2} [\Ccorr(\theta)]^2 \dderiv(\cos\theta)
  = \sum_{\ell=2}^{\ellmax} \Cl I_{\ell\ell'} \Cee_{\ell'}.
\end{equation}
As discussed above, this definition applies to both full-sky and cut-sky
maps.  For the case of full-sky maps the full-sky $\Cl$ from
Eq.~(\ref{eq:Cl-full-sky}) are used in the sum on the right-hand side,
whereas for the case of cut-sky maps the cut-sky $\pseudoClest$ from
Eq.~(\ref{eq:Ctheta-cut-sky-expansion}) are used.  Throughout we will either refer to the
$\Shalf$ statistic generically or make it clear the context in which it is
calculated.  This statistic has not been optimized in any way, except
crudely by the choice of the limits of integration, particularly the upper
one which has been chosen to be a convenient value.  We consistently resist
the temptation to optimize these limits in order to minimize the
oft-repeated criticism that the statistic is \textit{a posteriori}.  In
acknowledgement and partial response to that objection, we note that the
statistical significance of the absence of large-angle correlations is not
particularly dependent either on the precise value of either limit (so long
as the range of integration focuses on large scales) nor on the particular
choice of reasonable integrand.\footnote{In another paper, looking at the
  predictions for the two-point angular correlation function of temperature
  with polarization, specifically the $Q$ Stokes parameter \citep{CHSS-TQ},
  we optimized the upper and lower limits of integration, and considered
  both $[\Ccorr^{TQ}(\theta)]^2$ and $\Ccorr^{TQ}(\theta)$ as integrands in
  the equivalent of (\ref{eq:Shalf}).  However, in that case we were
  \textit{a priori} optimizing a statistic for a specific purpose --
  differentiating between two models.  Furthermore, it was found
  that replacing $[\Ccorr^{TQ}(\theta)]^2$ in the integrand with
  $\Ccorr^{TQ}(\theta)$ makes no qualitative difference in the
  conclusions.}
  
The sum in equation~(\ref{eq:Shalf}) shows how to quickly and easily
calculate $\Shalf$ in terms of the $\Cl$ or $\pseudoClest$ from the
Legendre series~(\ref{eq:Ctheta-full-sky-expansion}) or
(\ref{eq:Ctheta-cut-sky-expansion}).  The ${I}_{\ell \ell'}$ are the
components of a matrix of integrals over products of Legendre polynomials
and are simply related to the $\mathcal{I}_{\ell\ell'}(1/2)$ calculated in
Appendix A of \cite{CHSS-WMAP5} by ${(4\upi)^2} I_{\ell\ell'} =
{(2\ell+1)(2\ell'+1)} \mathcal{I}_{\ell\ell'}(1/2)$.  The sum in the
$\Shalf$ expression~(\ref{eq:Shalf}) ranges from $\ell=2$ to
$\ell=\ellmax$.  The lower limit is due to the monopole and dipole being
removed from the map.  We remove the monopole both because its amplitude is
significantly larger than those of the other multipoles and because we are
interested in the correlations among fluctuations not in the background
value.  We remove the entire dipole because it is dominated by the Doppler
dipole -- the (uninteresting) contribution due to our peculiar motion
through the Universe; this is approximately two orders of magnitude larger
than the expected underlying dipole in the CMB rest frame.  Once it is
possible to measure the Doppler contribution to better than $1\%$, it will
be far preferable to remove the Doppler dipole, and set $\ell=1$ as the
lower limit of the sum in expression~(\ref{eq:Shalf}).  For the upper limit
there is some freedom in the choice of $\ellmax$.  Since $\Cl\sim
\ell^{-2}$ we would expect that the result is independent of our choice
provided that $\ellmax$ is `large enough'.  However, since we will find
small values of $\Shalf$, the exact choice does have a slight effect on the
final values.  We have consistently chosen $\ellmax=100$ for all
calculations of $\Shalf$ in this work.  The effect of this choice on the
value of $\Shalf$ depends on the map and mask employed but is always less
than one per cent (for $\ellmax \geq 100$).  For example, using a larger
value of $\ellmax$ with the \Planck\ \smica\ full-sky map the maximum
deviation in $\Shalf$ is about $0.08$ per cent. For this map and the U74
mask the maximum deviation is about $0.4$ per cent.  Larger variations in
the $\Shalf$ value occur between the maps as seen in
Tables~\ref{tab:S12-results} and~\ref{tab:S12-results-DQ}.  For example, in
Table~\ref{tab:S12-results} there is a $0.7$ per cent change between the
$\Shalf$ value for the \Planck\ \smica\ and \nilc\ maps with the U74 mask,
however, even this difference only corresponds to a change in the third
digit in the $p$-value (from $0.191$ to $0.195$).  In practice, the choice
of $\ellmax$ should be used consistently in analysing both the data and
realizations which will further mitigate its effect.

The calculation of $\Shalf$ has therefore been reduced to finding the
angular power spectrum either over the full-sky or cut-sky for
some map of the CMB temperature.  However, a number of important choices
must be made, which we now discuss.

First, there are a number of maps available for analysis.  In each data
release, the \WMAP\ team included individual band maps and a full-sky
Independent Linear Combination (ILC) map designed to be as close to the
foreground-subtracted CMB as possible.  The \Planck\ team released
individual band maps and three different foreground-subtracted maps --
\nilc, \sevem, \smica\ -- in their initial 2013 release, although they had
many more, including one they called the \CR\ map\footnote{The \CR\ map was
  subsequently released after this study was completed.}.  Here we will
analyse the seven and nine-year \WMAP\ $V$ and $W$ band maps and the ILC
map, the \Planck\ High Frequency Instrument (\HFI) $100\unit{GHz}$ and Low
Frequency Instrument (\LFI) $70\unit{GHz}$ maps (these two channels are
expected to be least contaminated by foregrounds), and its \nilc, \sevem,
and \smica\ maps.\footnote{All CMB data is available from the Lambda site,
  \url{http://lambda.gsfc.nasa.gov/}, including links to both \WMAP\ and
  \Planck\ results.  The \Planck\ results may directly be obtained via the
  \Planck\ Legacy Archive, \url{http://archives.esac.esa.int/pla/}.}

Once we have a map, we must also choose the resolution of the maps to be
analysed.  A higher resolution will minimize resolution-dependent effects.
On the other hand, to reduce the computation time, particularly when
generating statistics from realizations of \LCDM, a low resolution is
preferred.  As a compromise we have chosen the \healpix\footnote{The
  \healpix\ source code is freely available from
  \url{healpix.sourceforge.net}.} resolution $\Nside=128$ for all studies
in this work.

To work at $\Nside=128$ we degrade the high resolution maps by averaging
over pixels using \texttt{ud\_grade} from \healpix.  This process follows
that used for degrading the masks discussed below.  To gain a computational
advantage from working at lower resolution, realizations are generated at
$\Nside=128$ directly.  Further, they are only generated including modes up
to $\ellmax=100$.  It has been verified that neither degrading from higher
resolutions realizations nor increasing $\ellmax$ affects the final
results.  This is not surprising given that $\Shalf$ is only weakly
dependent on small scale and large-$\ell$ behaviour.

Even with the existence of cleaned, full-sky maps the concern of residual
contamination, particularly on the largest angular scales, remains.  For
this reason it is desirable to remove the most contaminated regions of the
sky and only analyse the cleanest ones.  The \Planck\ analysis used the U73
mask which leaves a sky fraction $\fsky=0.73$ \citep{Planck-R1-XXIII}.
This mask is not publicly available but is constructed from the union of
the validity masks provided with the full-sky maps \citep{Planck-R1-XII}.
For the \nilc, \sevem, and \smica\ maps these masks are available.  For the
\CR\ map only a minimal version of the mask is provided.  Taking the union
of these four masks produces what we call the U74 mask, a close
approximation of the U73 mask but with $\fsky=0.74$.  For \WMAP\ we use
their extended temperature mask from their nine-year data release, named
KQ75y9, which has $\fsky=0.69$.

These masks are provided at high resolution; $\Nside=2048$ for the U74 mask
and $\Nside=1024$ for the KQ75y9 mask.  To degrade the masks to our working
resolution of $\Nside=128$ we follow the prescription defined in
\cite{Planck-R1-XXIII}: first the mask is degraded to $\Nside=128$ using
\texttt{ud\_grade} from \healpix, then any pixel with a value less then
$0.8$ is set to zero, otherwise it is set to one.  With this prescription
the $\Nside=128$ masks have sky fractions of $\fsky=0.72$ for U74 and
$\fsky=0.67$ for KQ75y9.

\begin{figure}
  \includegraphics[width = \linewidth]{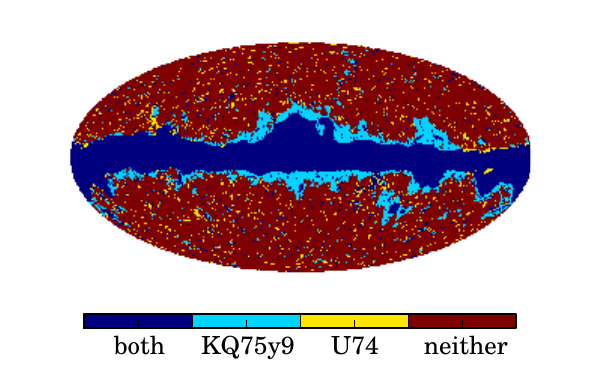}
  \caption{Masks used in this work. A pixel may be removed by both masks
    (dark blue), only the KQ75y9 mask (light blue), only the U74 mask
    (yellow), or by neither mask (red).}
  \label{fig:masks-comparison}
\end{figure}

Despite the KQ75y9 mask removing more pixels, the U74 mask is not fully
contained within it.  A comparison of the two masks is given in
Fig.~\ref{fig:masks-comparison}.  As can be seen, the two masks mostly
coincide, though there are many small regions of pixels only contained in
one of the two masks.  In particular there are pixels that are excluded by
the U74 mask but included by the KQ75y9 mask and the KQ75y9 mask generally
removes more of the region around the Galactic centre than the U74 mask.
These small differences have a noticeable effect on the calculated cut-sky
$\Shalf$\@.

It is important that comparisons of data and simulations are made
consistently.  In addition to the choices discussed above, cut-sky data
will always be compared to cut-sky realizations, with the maps in all cases
treated as similarly as possible.  This is particularly important since, as
noted above, for cut skies the cut-sky $\Cl$ are employed in the calculation
of $\Shalf$\@.  In this work we are \emph{not} interested in reconstructing
the full-sky angular correlations. Instead, we find that angular
correlations on the cut-sky are unusually low. We thus do not make
statements about the full-sky CMB, which at any rate cannot be reliably
observed, and for which a maximum-likelihood estimator may be more
appropriate \citep{Efstathiou2004-MLE, Efstathiou2010, Pontzen2010}.  Even so,
reconstructing the full-sky from a cut-sky requires extra assumptions and
may introduce its own biases \citep{CHSS2011}.

Extracting the $\Cl$ from a map, particularly from a masked map, also
requires some care.  We use \spice\ \citep{polspice} for this purpose since
it calculates the $\pseudoClest$ which appear in the Legendre
series~(\ref{eq:Ctheta-cut-sky-expansion}).  For cut skies there is the
added issue that, even if the full-sky does not include a monopole or
dipole, these modes will exist in the portion of the sky included for
evaluation.  If we knew that the full-sky map did not contain a residual
monopole or dipole, then we could proceed without further concern.
Unfortunately, with real data this is not known, particularly for
individual frequency band maps which definitely have Galactic
contamination.  We therefore remove the average monopole and dipole from
all maps prior to extracting the $\Cl$\@.  For the monopole, we do this by
subtracting the average value of the temperature over the portion of the
sky that is being retained; for the dipole we find the best-fitting dipole
over the retained sky and subtract that dipole. (In \spice\ this removal is
a built-in feature which we employ in our analysis.)  When analysing a
cut-sky, this procedure generically introduces a monopole and dipole (and
alters the other multiples) into the equivalent full-sky map.  Though this
may seem to be a problem, again recall that the cut-sky analysis is
self-contained and internally consistent since the data and realizations
are treated identically. The cut-sky statistics are \emph{not} estimators
of the full-sky, as again made clear by this monopole and dipole removal.

There is also the question of the effect of our motion with respect to the
CMB rest frame on the quadrupole.  Just as that motion, with velocity
$\beta\equiv v/c \sim 10^{-3}$, induces a dipole with amplitude
$\mathcal{O}\left(\beta\right)$ times the monopole, it also induces a
Doppler quadrupole (DQ) with amplitude $\mathcal{O}\left(\beta^2\right)$
times the monopole.  The naive expectation that since $\beta^2 \sim
10^{-6}$ the DQ will be an unimportant contribution to the cosmological
quadrupole is not obviously true at least in part because the measured
quadrupole is much smaller than the theoretical expectation.  For each map
we analyse both the DQ uncorrected and the DQ corrected map to gauge the
importance of this effect.  The one exception is the
\Planck\ \LFI\ $70\unit{GHz}$ map, where (at least part of) the DQ has been
accounted for in the calibration procedure.  See
\citet{Planck-R1-V,CHSS-Planck-R1-alignments} for a more detailed
discussion of this issue.

The effect of the boosted black body DQ correction on $\Shalf$ is shown in
Tables~\ref{tab:S12-results} and \ref{tab:S12-results-DQ} where it is found
that the effect on $p$-values is much less than the differences among the
\Planck\ maps, and is thus not significant.  The DQ correction is
frequency-dependent, and subsequent to this analysis
\citet{Planck-R1-XXIII} was updated to include estimates of DQ correction
factors for each of their released combined maps.  Since neither a complete
description of how these correction factors were calculated nor all the
data required to calculate such correction factors were made publicly
available, and since the effect on $\Shalf$ was negligible from the simple
estimate of the DQ correction, these correction factors have not been
included in our analysis.

\section{Results}

\begin{figure}
  \includegraphics[width = \linewidth]{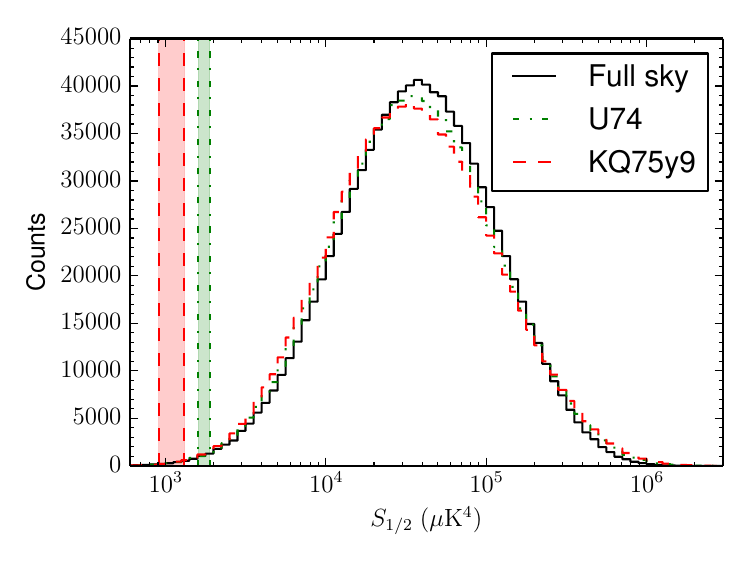}
  \caption{Distribution of $\Shalf$ values from $10^6$ realizations of the
    best-fitting \LCDM\ model for full and masked skies. The shaded regions
    (green, dash-dotted for the U74 mask and red, dashed for the KQ75y9
    mask) represent the spread of the the observed values as given in
    Tables~\ref{tab:S12-results} and \ref{tab:S12-results-DQ}.  Masking
    only slightly affects the expected distributions and the observations
    are in the small $\Shalf$ tail of the distribution for both masks
    considered in this work.}
  \label{fig:S12-histograms}
\end{figure}

\begin{table}
   \caption{Smallness of $\Shalf$ for maps without the DQ correction. We
     analyse the cleaned maps from \Planck: \nilc, \sevem, and \smica, as
     well as from \WMAP: seven and nine-year ILC.  We also analyse the
     individual frequency band maps from \Planck: \HFI\ $100\unit{GHz}$ and
     \LFI\ $70\unit{GHz}$, as well as from \WMAP: seven and nine-year $W$
     and $V$ bands.  For each map, we use both the U74 and KQ75y9 masks.
     In all cases residual monopole and dipole contributions have been
     subtracted from the map after masking.  For each map and mask we
     report the $\Shalf$ value and the associated $p$-value -- the fraction
     of realizations of \LCDM\ in the \Planck\ best-fitting \LCDM\ model
     with an $\Shalf$ no larger than the reported value.}
  \label{tab:S12-results}
\begin{tabular}{ld{4.1}d{3}d{4.1}d{3}} \hline
    & \multicolumn{2}{c}{U74}
    & \multicolumn{2}{c}{KQ75y9}
    \\
    \multicolumn{1}{c}{Map}
    & \multicolumn{1}{c}{$\Shalf\unit{(\muK)^4}$}
    & \multicolumn{1}{c}{$p$ (\%)}
    & \multicolumn{1}{c}{$\Shalf\unit{(\muK)^4}$}
    & \multicolumn{1}{c}{$p$ (\%)}
    \\ \hline
    \WMAP\ ILC 7yr & 1582.3 & 0.193 	& 1225.8 & 0.085
    \\
    \WMAP\ ILC 9yr & 1626.0 & 0.211	& 1278.2 & 0.100
    \\
    \Planck\ \smica & 1577.7 & 0.191 	& 1022.3 & 0.044
    \\
    \Planck\ \nilc & 1589.3 & 0.195	& 1038.2 & 0.047
    \\
    \Planck\ \sevem & 1657.7 & 0.225	& 1153.4 & 0.069
    \\
    \\
    \WMAP\ $W$ 7yr & 1863.6 & 0.316	& 1133.9 & 0.065
    \\
    \WMAP\ $W$ 9yr & 1887.1 & 0.329 	& 1142.6 & 0.068
    \\
    \Planck\ \HFI\ $100$ & 1682.1& 0.235 & 911.6 & 0.027
    \\
    \\
    \WMAP\ $V$ 7yr & 1845.0 & 0.307 	& 1290.9 & 0.104
    \\
    \WMAP\ $V$ 9yr & 1850.0 & 0.309	& 1281.8 & 0.101
    \\
    \Planck\ \LFI\ $70^a$ & \multicolumn{1}{c}{---} &
    \multicolumn{1}{c}{---} & \multicolumn{1}{c}{---} &
    \multicolumn{1}{c}{---}
    \\
    \hline
  \end{tabular}
\\ ${}^a$The calibration of the \Planck\ \LFI\ $70\unit{GHz}$ channel
includes the DQ correction.  See
\citet{Planck-R1-V,CHSS-Planck-R1-alignments} for details.
\end{table} 

\begin{table}
  \caption{Same as Table~\ref{tab:S12-results} now with the DQ
    corrected maps.}
  \label{tab:S12-results-DQ}
\begin{tabular}{ld{4.1}d{3}d{4.1}d{3}} \hline
    & \multicolumn{2}{c}{U74}
    & \multicolumn{2}{c}{KQ75y9}
    \\
    \multicolumn{1}{c}{Map}
    & \multicolumn{1}{c}{$\Shalf\unit{(\muK)^4}$}
    & \multicolumn{1}{c}{$p$ (\%)}
    & \multicolumn{1}{c}{$\Shalf\unit{(\muK)^4}$}
    & \multicolumn{1}{c}{$p$ (\%)}
    \\ \hline
    \WMAP\ ILC 7yr	& 1620.3 & 0.208 	& 1247.0 & 0.090
    \\ 
    \WMAP\ ILC 9yr 	& 1677.5 & 0.232 	& 1311.8 & 0.109
    \\
    \Planck\ \smica	& 1606.3 & 0.202	& 1075.5 & 0.053
    \\
    \Planck\ \nilc	& 1618.6 & 0.208 	& 1096.2 & 0.058
    \\
    \Planck\ \sevem	& 1692.4 & 0.239 	& 1210.5 & 0.082
    \\
    \\
    \WMAP\ $W$ 7yr  	& 1839.0 & 0.304	& 1128.5 & 0.064
    \\
    \WMAP\ $W$ 9yr	& 1864.2 & 0.317	& 1138.3 & 0.066
    \\ 
    \Planck\ \HFI\ $100$ & 1707.5 & 0.245	& 916.3 & 0.028
    \\
    \\
    \WMAP\ $V$ 7yr 	& 1829.2 & 0.300	& 1276.2 & 0.099
    \\
    \WMAP\ $V$ 9yr 	& 1840.4 & 0.304	& 1268.8 & 0.097
    \\
    \Planck\ \LFI\ $70$	& 1801.7 & 0.287	& 1282.1 & 0.101
    \\
    \hline
  \end{tabular}
\end{table} 

Histograms of $\Shalf$ values from $10^6$ realizations of the
\Planck\ best-fitting \LCDM\ model (based on their temperature only data)
are shown in Fig.~\ref{fig:S12-histograms}.  Included in the figure are the
full-sky and cut-sky $\Shalf$.  As seen in the figure, masking has a small
effect; the peak of the distribution is shifted to slightly smaller values
due to masking, but this does not have a noticeable change on the tail of
the distribution.  Regardless, in comparing cut-sky $\Shalf$ between the
data and our realizations, we always compare the one set of cut-sky data to
the same set of cut-sky realizations.

The $\Shalf$ values for the various map and mask combinations are given in
Table~\ref{tab:S12-results} for the case when the maps are not DQ corrected
and in Table~\ref{tab:S12-results-DQ} when the DQ correction has been
applied.  As discussed above, the realization maps are treated precisely
like the data maps -- they are masked, then monopole and dipole are
subtracted before $\Shalf$ is computed.  Given that the value of $\Shalf$
on masked skies is extremely low compared to the typical value, having
$10^6$ is necessary to make quantitatively precise statements.  For each
computed value of $\Shalf$ reported, we also report the $p$-value -- the
fraction of realizations (expressed in per cent) that have an $\Shalf$ at
least as low.  This we interpret as the probability of obtaining a value of
$\Shalf$ this low by random chance in the best-fitting model of \LCDM.

An alternative approach is to allow for variations of the best-fitting
parameters within their error bars, for example by examining a Monte Carlo
Markov chain of the parameters rather than just performing realizations of
the best-fitting values.  (Such an approach was taken for example in
\cite{CHSS-WMAP5}.)  This will affect the results only weakly, because
varying the parameters within their error bars will cause the expected
low-$\ell$ $\Cl$ to vary by much less than their cosmic variance.

\begin{figure}
  \includegraphics[width = \linewidth]{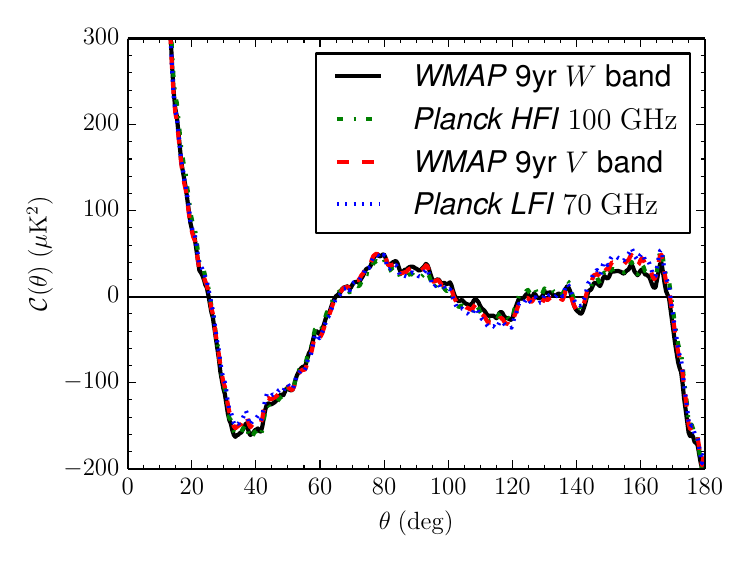}
  \caption{Cut-sky $\Ccorr(\theta)$ using the KQ75y9 mask for individual
    frequency band maps.  Shown are correlation functions from the
    \WMAP\ nine-year $W$ (black, solid line) and $V$ (red, dashed line)
    bands along with the \Planck\ \HFI\ $100\unit{GHz}$ (green, dash-dotted
    line) and \LFI\ $70\unit{GHz}$ (blue, dotted line) maps.  The curves
    for the \WMAP\ seven-year band maps are nearly identical to those from
    the nine-year maps and are not included for clarity.  In all cases the
    correlation functions are in excellent agreement across the data
    releases and frequency bands.  (Note the range on $y$-axis has been
    greatly reduced as compared to Fig.~\ref{fig:Ctheta-smica} to allow for
    \emph{any} difference to be noticeable by eye.)}
  \label{fig:Ctheta-bands}
\end{figure}

The cut-sky $\Shalf$ values presented in Tables~\ref{tab:S12-results} and
\ref{tab:S12-results-DQ} show that the region outside the masks is
consistently observed and cleaned in all the data releases mostly
independent of analysis procedures.  In Fig.~\ref{fig:Ctheta-bands} we plot
$\Ccorr(\theta)$ for the \WMAP\ nine-year $V$ and $W$ bands with the KQ75y9
mask and the \Planck\ \HFI\ $100\unit{GHz}$ and \LFI\ $70\unit{GHz}$ bands
also with the KQ75y9 mask.  One can see that the cut-sky angular
correlation functions are remarkably consistent across instruments and
wavebands.  (And also across \WMAP\ data releases.  We have chosen not to
plot the \WMAP\ seven-year correlation functions because they are nearly
indistinguishable from the nine-year functions.)  We can thus place great
confidence in the cut-sky $\Shalf$ results derived from \WMAP\ and \Planck.
These results can be summarized as follows.
\begin{itemize}
\item Regardless of the maps, the cut-sky $\Shalf$ is very low, with
  $p$-values ranging from $0.027$ per cent for the
  \Planck\ \HFI\ $100\unit{GHz}$ map with the KQ75y9 mask to 
  $0.329$ per cent for the \WMAP\ nine-year $W$ band map with the U74
  mask.
\item The cleaned maps have a smaller variation in $\Shalf$ values with a
  $p$-value always less than about $0.239$ per cent for the U74 mask and
  less than about $0.109$ per cent for the KQ75y9 mask.
\item The \Planck\ maps typically have smaller $\Shalf$ values than the
  \WMAP\ maps. (The two slight exceptions are the DQ corrected
  \Planck\ \LFI\ $70\unit{GHz}$ band with the KQ75y9 mask and the
  \Planck\ \sevem\ map for the U74 mask.)
\item The only clear systematic trend is that the KQ75y9 mask consistently
  yields a lower cut-sky $\Shalf$ than does the U74 mask.  Presumably this
  is due to the larger region around the Galactic centre excluded by the KQ75y9
  mask (see Fig \ref{fig:masks-comparison}).
\item The DQ correction has little effect, in most cases tending to
  slightly increase $\Shalf$ in the \Planck\ maps and decrease it in the
  \WMAP\ ones.  This is in contrast to the importance of applying the DQ
  correction for full-sky alignment studies
  \citep{CHSS-Planck-R1-alignments}.
\end{itemize}

Overall, the data very consistently show a lack of correlations on large
angular scales outside the Galactic region (as defined by the two masks
employed).  The $p$-value for the $\Shalf$ statistic has remained small and
of comparable size throughout the \WMAP\ data releases and now with the
first \Planck\ results.  This is remarkable given the improvements in
statistics, cleaning, beams, masks, and other systematics.  Further, this
is in contrast to the full-sky $\Shalf$ which vary significantly from data
release to data release and from map to map.  The behaviour of the full-sky
$\Shalf$ is discussed in more detail in \citet{CHSS-Planck-R1-alignments}.
It suffices here to note that the full-sky value of $\Shalf$ varies from a
low of $3766\unit{(\muK^4)}$, from the \Planck\ \sevem\ map, to a high of
$8938\unit{(\muK^4)}$, calculated from the seven-year \WMAP\ reported values
of the angular power spectrum based on a maximum likelihood estimator.

We again emphasize that the two-point angular correlation function that we
have calculated is monopole and dipole-subtracted.  However, once the
Doppler dipole is sufficiently well determined, only it should be removed
and the underlying cosmological contribution to the dipole retained in
$\Ccorr(\theta)$ and thus in $\Shalf$.

The measured lack of angular correlations in the dipole-subtracted sky has
an important consequence for the primordial dipole. If the missing
correlations are not a very unlikely fluke, nor (as our results indicate)
due to systematic errors or map-cleaning procedures, then they are caused
by some as-yet unidentified physical mechanism. It is difficult to see how
such a mechanism would set $\Ccorr(\theta)$ to be nearly zero on angular
scales greater than $60$ degrees when the primordial dipole is subtracted,
and yet somehow not also do so if the dipole were included.  Instead, for a
physical mechanism we would expect the total angular correlation function
including the contribution of the cosmological dipole to also be nearly
zero on these scales. In the best-fitting \LCDM\ model, the expected
contribution from the dipole alone is very large and generically spoils the
vanishing of $\Ccorr(\theta)$ on large angular scales.  Hence, if the
vanishing correlations are of cosmological origin, then the primordial
dipole is also expected to be very suppressed.

To be concrete, the expected value of $\Cee_1$ in the
best-fitting \LCDM\ model is approximately $3300\unit{\muK^2}$.  With this
value, the $\Cee_1^2$ contribution to $\Shalf$ in Eq.~(\ref{eq:Shalf}) alone
would contribute approximately $2.3\times10^5\unit{\muK^4}$ to $\Shalf$. (In
principle this could be compensated by the cross term, $\Cee_1\Cee_\ell$ with
$\ell\neq1$, which can be negative, however, in practice this does not occur
owing mainly to the smallness of $\Cee_2$.)  Roughly, for the $\Cee_1^2$
contribution to not make the $\Shalf$ `too large' the value of $\Cee_1$ must
also not be `too large'.  For example, requiring the contribution to $\Shalf$
to be comparable to current cut-sky values, that is a contribution of order
$1000\unit{\muK^4}$, places a limit $\Cee_1\la 200\unit{\muK^2}$. This has a
probability of occurring by chance in a realization of the best-fitting model
(due to cosmic variance) of less than approximately $0.4$ per cent.
Equivalently, to the extent that $\Cee_1$ contributions dominate the value of
$\Shalf$, in order to maintain a $p$-value for the $\Shalf$ less than $0.4$
per cent once the cosmological dipole is included requires that
$\Cee_1\la200\unit{\muK^2}$.  

To summarize, it seems unlikely that a physical mechanism
would predict that the $\Shalf$ calculated from a dipole-subtracted cut-sky
would be small but the $\Shalf$ calculated from a non-dipole-subtracted
cut-sky would not be.  This strongly suggests that if the lack of angular
correlations is physical in nature and not a statistical fluke, then a robust
prediction can be made that there is a very small cosmological dipole.  In a
future work we will develop this prediction more precisely.

\section{Conclusions}

The CMB shows a lack of correlations on large angular scales.  This can be
quantified by the $\Shalf$ statistics proposed by \citet{WMAP1-cosmology}
which is best calculated on the portion of the sky outside the Galaxy.
Unlike attempts to infer properties of the the full-sky correlation
function, the cut-sky $\Shalf$ appears remarkably robust and trustworthy.
In our analysis we find that the $p$-value for the observed cut-sky
$\Shalf$ in an ensemble of realizations of the best-fitting \LCDM\ model
never exceeds $0.33$ per cent for any of the analysed combinations of maps
and masks, with and without correcting for the Doppler quadrupole. This has
remained the case since the \WMAP\ three-year data release,\footnote{The
  one-year \WMAP\ release yielded slightly higher $p$-values -- $0.38$ per
  cent for the $V$ band, and $0.64$ per cent for the $W$ band
  \citep{CHSS-WMAP3, CHSS-WMAP5}.} for both the individual ($V$ and $W$)
band maps and the synthesized (ILC) map, and for the first \Planck\ data
release for both the \LFI\ and \HFI\ band maps and all the released
synthesized maps (\nilc, \smica, \sevem), when masked by either the
\WMAP\ KQ75y9 mask or the less conservative U74 mask (which is very similar
to the \Planck\ U73 mask).  The \HFI\ $100\unit{GHz}$ map -- the presumably
cleanest CMB band of \HFI -- with the more conservative mask that has been
defined by \WMAP\ gives a $p$-value of only $0.03$ per cent! As general
trends we note that a larger mask tends to produce smaller $p$-values, the
Doppler quadrupole correction does not change the results in a significant
way, and the \Planck\ data yield somewhat smaller $p$-values than the
\WMAP\ data.

This apparent lack of temperature correlations on large angular scales is
striking.  It is a robust observation that increases in statistical
significance from \COBE\ to \WMAP\ to \Planck. The consistency of the lack
of angular correlations greatly reduces the likelihood of instrumental
issues as a cause. Since all three missions observed the same sky, we could
be unlucky and live in a very atypical realization of the Universe. A
method of testing this hypothesis has been proposed that would utilize
the upcoming \Planck\ polarization data \citep{CHSS-TQ}.  If it is not a
statistical fluke and not an instrumental issue, it still could be caused
by foregrounds.  This appears also unlikely as the lack of correlations is
consistently seen in individual bands as well as in foreground cleaned
maps. Thus, to the best of our knowledge, the lack of angular correlations
is in contradiction with the idea of scale invariant, isotropic and
Gaussian perturbations seeded by cosmological inflation.

Attempts to explain this lack of correlations should also address the
various other anomalous aspects observed in the microwave sky. It turns out
that the lack of angular correlations puts very severe constraints on such
models. For example, a plausible explanation for the alignments of
multipole vectors or for the hemispherical asymmetry observed might have
been contamination by unaccounted foregrounds; however, one cannot easily
understand how a hypothetical foreground, which presumably should be
uncorrelated with the primordial temperature fluctuations, could cause an
almost exact cancellation of the primordial fluctuations at angular scales
above 60 degrees.

Several attempts have been made to explain the absence of large-angle
correlations as being due to an unknown foreground or, more generally, by
altering the procedure by which one arrives at the cleaned maps.  Indeed,
when the cleaned maps are altered in any way the microwave sky can easily
be made to appear less anomalous.   This
is not surprising; almost any random modification of the observed maps will
make them less anomalous. Though the removal of anomalies may be a
side effect of improved analysis procedures, using their removal as a
basis for judging the effectiveness of such a procedure is misguided. 

Finally, we emphasize that in order to be convincing, new theoretical
models to explain the observed large-angle anomalies must be based on the
statistics of realizations of that model, not just on having the mean
values of the model agree with observations.  In other words, CMB map
realizations based on the underlying new model should have $p$-values for
the measured statistics that are not unusually small.

The large-angle temperature-temperature correlations in the CMB outside the
Galaxy have been anomalously low in all relevant maps since the days of the
\COBEDMR.  The final \WMAP\ release and the initial \Planck\ release
confirm that anomaly.  After twenty years, we still await a satisfactory
explanation.

\section*{Acknowledgements}

We acknowledge valuable communications and discussions with F.~Bouchet,
C.~Burigana, J.~Dunkley, G.~Efstathiou, K.~Ganga, P.~Naselsky, H.~Peiris,
C.~R\"ath and D.~Scott.  GDS and CJC are supported by a grant from the US
Department of Energy to the Particle Astrophysics Theory Group at CWRU\@.
DH has been supported by the DOE, NSF, and the Michigan Center for
Theoretical Physics\@.  DJS is supported by the DFG grant RTG 1620 `Models
of gravity'. DH thanks the Kavli Institute for Theoretical Physics and GDS
thanks the Theory Unit at CERN for their respective hospitality.  This work
made extensive use of the \healpix{} package~\citep{healpix}.  The
numerical simulations were performed on the facilities provided by the Case
ITS High Performance Computing Cluster.  We acknowledge the use of the
Legacy Archive for Microwave Background Data Analysis (LAMBDA), part of the
High Energy Astrophysics Science Archive Center (HEASARC). HEASARC/LAMBDA
is a service of the Astrophysics Science Division at the NASA Goddard Space
Flight Center.  This paper made use of observations obtained with
\Planck\ (\url{http://www.esa.int/Planck}), an ESA science mission with
instruments and contributions directly funded by ESA Member States, NASA,
and Canada.

\bibliographystyle{mn2e_new}

\bibliography{planck_r1_ctheta}

\begin{thebibliography}{33}
\providecommand{\natexlab}[1]{#1}

\bibitem[{{Bennett} et~al.(2011)}]{WMAP7-anomalies}
{Bennett} C.L. et~al., 2011, {\apjs}, 192, 17

\bibitem[{{Bennett} et~al.(2013)}]{WMAP9-results}
{Bennett} C.L. et~al., 2013, \apjs, 208, 20

\bibitem[{{Chon} et~al.(2004){Chon}, {Challinor}, {Prunet}, {Hivon}, \&
  {Szapudi}}]{polspice}
{Chon} G., {Challinor} A., {Prunet} S., {Hivon} E., {Szapudi} I., 2004, \mnras,
  350, 914

\bibitem[{Copi et~al.(2007)Copi, Huterer, Schwarz, \& Starkman}]{CHSS-WMAP3}
Copi C.J., Huterer D., Schwarz D.J., Starkman G.D., 2007, \prd, 75, 023507

\bibitem[{{Copi} et~al.(2009){Copi}, {Huterer}, {Schwarz}, \&
  {Starkman}}]{CHSS-WMAP5}
{Copi} C.J., {Huterer} D., {Schwarz} D.J., {Starkman} G.D., 2009, \mnras, 399,
  295

\bibitem[{{Copi} et~al.(2010){Copi}, {Huterer}, {Schwarz}, \&
  {Starkman}}]{CHSS-review}
{Copi} C.J., {Huterer} D., {Schwarz} D.J., {Starkman} G.D., 2010, \advastro,
  2010, 78

\bibitem[{{Copi} et~al.(2011){Copi}, {Huterer}, {Schwarz}, \&
  {Starkman}}]{CHSS2011}
{Copi} C.J., {Huterer} D., {Schwarz} D.J., {Starkman} G.D., 2011, \mnras, 418,
  505

\bibitem[{{Copi} et~al.(2013){Copi}, {Huterer}, {Schwarz}, \&
  {Starkman}}]{CHSS-TQ}
{Copi} C.J., {Huterer} D., {Schwarz} D.J., {Starkman} G.D., 2013, \mnras, 434,
  3590

\bibitem[{{Copi} et~al.(2015){Copi}, {Huterer}, {Schwarz}, \&
  {Starkman}}]{CHSS-Planck-R1-alignments}
{Copi} C.J., {Huterer} D., {Schwarz} D.J., {Starkman} G.D., 2015, \mnras, 449,
  3458

\bibitem[{{Dup{\'e}} et~al.(2011){Dup{\'e}}, {Rassat}, {Starck}, \&
  {Fadili}}]{Dupe2011}
{Dup{\'e}} F.X., {Rassat} A., {Starck} J.L., {Fadili} M.J., 2011, \aap, 534,
  A51

\bibitem[{{Efstathiou}(2004)}]{Efstathiou2004-MLE}
{Efstathiou} G., 2004, \mnras, 348, 885

\bibitem[{{Efstathiou} et~al.(2010){Efstathiou}, {Ma}, \&
  {Hanson}}]{Efstathiou2010}
{Efstathiou} G., {Ma} Y., {Hanson} D., 2010, \mnras, 407, 2530

\bibitem[{{Francis} \& {Peacock}(2010)}]{Francis2010}
{Francis} C.L., {Peacock} J.A., 2010, \mnras, 406, 14

\bibitem[{{G{\'o}rski} et~al.(2005){G{\'o}rski}, {Hivon}, {Banday}, {Wandelt},
  {Hansen}, {Reinecke}, \& {Bartelmann}}]{healpix}
{G{\'o}rski} K.M., {Hivon} E., {Banday} A.J., {Wandelt} B.D., {Hansen} F.K.,
  {Reinecke} M., {Bartelmann} M., 2005, \apj, 622, 759

\bibitem[{{Gruppuso}(2014)}]{Gruppuso2014}
{Gruppuso} A., 2014, \mnras, 437, 2076

\bibitem[{{Hajian}(2007)}]{Hajian2007}
{Hajian} A., 2007, {astro-ph/0702723}

\bibitem[{{Hauser} \& {Peebles}(1973)}]{Hauser1973}
{Hauser} M.G., {Peebles} P.J.E., 1973, \apj, 185, 757

\bibitem[{{Hinshaw} et~al.(1996){Hinshaw}, {Branday}, {Bennett}, {G{\'o}rski},
  {Kogut}, {Lineweaver}, {Smoot}, \& {Wright}}]{DMR4-Ctheta}
{Hinshaw} G., {Branday} A.J., {Bennett} C.L., {G{\'o}rski} K.M., {Kogut} A.,
  {Lineweaver} C.H., {Smoot} G.F., {Wright} E.L., 1996, \apj, 464, L25

\bibitem[{{Kim} \& {Naselsky}(2011)}]{Kim2011}
{Kim} J., {Naselsky} P., 2011, \apj, 739, 79

\bibitem[{{Planck Collaboration I}(2014)}]{Planck-R1-I}
{Planck Collaboration I}, 2014, \aap, 571, A1

\bibitem[{{Planck Collaboration V}(2014)}]{Planck-R1-V}
{Planck Collaboration V}, 2014, \aap, 571, A5

\bibitem[{{Planck Collaboration XII}(2014)}]{Planck-R1-XII}
{Planck Collaboration XII}, 2014, \aap, 571, A12

\bibitem[{{Planck Collaboration XV}(2014)}]{Planck-R1-XV}
{Planck Collaboration XV}, 2014, \aap, 571, A15

\bibitem[{{Planck Collaboration XVI}(2014)}]{Planck-R1-XVI}
{Planck Collaboration XVI}, 2014, \aap, 571, A16

\bibitem[{{Planck Collaboration XVII}(2014)}]{Planck-R1-XVII}
{Planck Collaboration XVII}, 2014, \aap, 571, A17

\bibitem[{{Planck Collaboration XXIII}(2014)}]{Planck-R1-XXIII}
{Planck Collaboration XXIII}, 2014, \aap, 571, A23

\bibitem[{{Planck Collaboration XXIV}(2014)}]{Planck-R1-XXIV}
{Planck Collaboration XXIV}, 2014, \aap, 571, A24

\bibitem[{{Pontzen} \& {Peiris}(2010)}]{Pontzen2010}
{Pontzen} A., {Peiris} H.V., 2010, \prd, 81, 103008

\bibitem[{{Raki{\'c}} et~al.(2006){Raki{\'c}}, {R{\"a}s{\"a}nen}, \&
  {Schwarz}}]{Rakic2006a}
{Raki{\'c}} A., {R{\"a}s{\"a}nen} S., {Schwarz} D.J., 2006, \mnras, 369, L27

\bibitem[{{Rassat} \& {Starck}(2013)}]{Rassat2013}
{Rassat} A., {Starck} J.L., 2013, \aap, 557, L1

\bibitem[{{Sarkar} et~al.(2011){Sarkar}, {Huterer}, {Copi}, {Starkman}, \&
  {Schwarz}}]{SHCSS2011}
{Sarkar} D., {Huterer} D., {Copi} C.J., {Starkman} G.D., {Schwarz} D.J., 2011,
  Astropart. Phys., 34, 591

\bibitem[{{Spergel} et~al.(2003)}]{WMAP1-cosmology}
{Spergel} D.N. et~al., 2003, \apjs, 148, 175

\bibitem[{{Zhang}(2012)}]{Zhang2012}
{Zhang} S., 2012, \apjl, 748, L20

\end{thebibliography}

\bsp

\label{lastpage}

\end{document}